\documentclass{article}
\usepackage{caption} % For controlling table captions
\usepackage{pifont}
\usepackage{enumitem}

\usepackage{microtype}
\usepackage{graphicx}
\usepackage{subfigure}
\usepackage{booktabs} % for professional tables
\usepackage{bbding} % for checkmark
\usepackage{adjustbox}
\usepackage{multirow}
\usepackage{makecell}
\usepackage{hyperref}
\usepackage{xcolor}
\newcommand{\fundwo}[1]{\textcolor{black}{#1}}
\newcommand{\yang}[1]{\textcolor{black}{#1}}
\newcommand{\slwu}[1]{\textcolor{black}{#1}}
\usepackage[bottom]{footmisc}

\usepackage[accepted]{icml2025}

\usepackage{amsmath}
\usepackage{amssymb}
\usepackage{mathtools}
\usepackage{amsthm}

\usepackage[capitalize,noabbrev]{cleveref}

\theoremstyle{plain}

\theoremstyle{definition}

\theoremstyle{remark}

\usepackage[textsize=tiny]{todonotes}

\icmltitlerunning{MuseControlLite: Multifunctional Music Generation with Lightweight Conditioners}

\begin{document}

\twocolumn[
\icmltitle{MuseControlLite: Multifunctional Music Generation \\ with Lightweight Conditioners}



\icmlsetsymbol{equal}{*}
\begin{icmlauthorlist}
\icmlauthor{Fang-Duo Tsai}{NTU}
\icmlauthor{Shih-Lun Wu}{MIT}
\icmlauthor{Weijaw Lee}{NTU}
\icmlauthor{Sheng-Ping Yang}{NTU}
\icmlauthor{Bo-Rui Chen}{NTU}
\icmlauthor{Hao-Chung Cheng}{NTU}
\icmlauthor{Yi-Hsuan Yang}{NTU}
%\icmlauthor{Firstname8 Lastname8}{yyy,comp}
\end{icmlauthorlist}

%Department, University, City, Region, Country
\icmlaffiliation{NTU}{National Taiwan University, Taipei, Taiwan.}
\icmlaffiliation{MIT}{Massachusetts Institute of Technology, Cambridge, MA, United States}

%\icmlaffiliation{NTU}{National Taiwan University}
%\icmlaffiliation{MIT}{Massachusetts Institute of Technology}
\icmlcorrespondingauthor{Fang-Duo Tsai}{fundwotsai2001@gmail.com}

\icmlkeywords{Machine Learning, ICML}

\vskip 0.3in
]



\printAffiliationsAndNotice{} % 

\begin{abstract}
We propose MuseControlLite, a lightweight mechanism designed to fine-tune text-to-music generation models for precise conditioning using various time-varying musical attributes and reference audio signals.
The key finding is that positional embeddings, which have been seldom used by text-to-music generation models in the conditioner for text conditions, are critical when the condition of interest is a function of time. Using melody control as an example, our experiments show that simply adding rotary positional embeddings to the decoupled cross-attention layers increases control accuracy from 56.6\% to 61.1\%, while requiring 6.75 times fewer trainable parameters than state-of-the-art fine-tuning mechanisms, using the same pre-trained diffusion Transformer model of Stable Audio Open. We evaluate various forms of musical attribute control, audio inpainting, and audio outpainting, demonstrating improved controllability over MusicGen-Large and Stable Audio Open ControlNet at a significantly lower fine-tuning cost, \yang{with only 85M trainable parameters}.
Source code, model checkpoints, and demo examples are available at: \url{https://MuseControlLite.github.io/web/}.
\end{abstract}

\section{Introduction}
\label{submission}

Text-to-music generation models have recently gained popularity as they hold the promise of empowering everyone to create high-quality and expressive music without much musical training and a reduced time cost~\cite{musicgen}. However, for people who desire to be more deeply involved in the creation process itself rather than the final output only, mechanisms to exert controllability going beyond the simple text prompt have been considered critical. 
As such, recent months have seen an increasing body of research concerning the addition of fine-grained, time-varying control to text-to-music generation, such as those related to musical aspects of chords, rhythm, melody, and dynamics. Exemplars include Music ControlNet~\cite{musiccontrolnet}, MusiConGen~\cite{musicongen}, JASCO~\cite{tal2024joint}, and DITTO~\cite{novack2024ditto}, to name just a few.
%evolved, it became clear that high-level textual descriptions alone are insufficient. These works showed that AI-generated music can also be conditioned on musical attributes like chords, rhythm, and melody, enabling more fine-grained control. 

Despite the exciting progress that has been made, we find two avenues for improvement. 
First, existing models might be over-parameterized. 
In view of the success of ControlNet~\cite{zhang2023adding, zhao2024uni} in adding spatial control for text-to-image generation,
%Drawing inspirations from the neighboring field of text-to-image generation, 
%spatial control techniques in 
a prominent approach 
%in the music domain 
has been to use similar idea
%the idea of ControlNet 
to fine-tune pre-trained text-to-music models to add conditioners for time-varying conditions. 
For example, treating mel-spectrograms as images, Music ControlNet~\cite{musiccontrolnet} offers multiple conditions of melody, rhythm and dynamics. 
%using latent diffusion instead and by 
%and audio quality by 
%uses extracted or constructed time-varying signals to guide music generation in the mel-spectrogram space. More recently, \citet{hou2024editing} 
%Follow-up research
Stable Audio Open ControlNet~\cite{hou2024editing} further improves audio quality with latent diffusion, adapting the original U-Net-based  ControlNet encoder to a diffusion Transformer architecture~\cite{peebles2023scalable}. %and introduced zero-initialized linear layers to bridge ControlNet with the Stable Audio Open\cite{evans2024stable} pretrained backbone, demonstrating melody-conditioned generation as a form of music editing.
However, the ControlNet approach requires duplicating half of the diffusion model as a trainable copy~\cite{zhang2023adding}, leading to increased training and inference times. 
%We find this approach to be over-parameterized. 
Lighter alternatives for fine-tuning, such as the idea of \emph{decoupled cross-attention} presented in IP-adapter~\cite{ye2023ip}, can be studied.

\yang{Second, while fine-tuning a text-to-music generation model to consider conditions of musical attributes or conditions of reference audio signals \cite{tsai2024audio} have been studied separately, little work has been done to consider both types of conditions at the same time. 
This capability may enable various creative use cases, such as adding melody conditions while performing audio inpainting or outpainting.}

%\fundwo{Second, while controllable music generation has received considerable attention, audio inpainting and outpainting conditioned on musical attributes remain underexplored. This capability could be particularly useful when the desired audio length exceeds the diffusion model's limit.}
%\fundwo{Second, while M} 
%We propose MuseControlLite, a lightweight fine-tuning mechanism with fewer trainable parameters yet empirically more accurate control of time-varying conditions than existing approaches, for controllable text-to-music generation.

\fundwo{We propose MuseControlLite, a lightweight fine-tuning mechanism with significantly fewer trainable parameters
%, yet empirically more accurate control over time-varying conditions than existing approaches, 
for controllable text-to-music generation.}
\fundwo{The key finding is that the decoupled cross-attention mechanism~\cite{ye2023ip}, when augmented with positional embeddings, achieves superior performance while using nearly an order of magnitude fewer trainable parameters than ControlNet-based mechanisms (85M vs. 572M).}
%but it needs \emph{positional embeddings} to work well.
%Interestingly, our pilot study shows that decoupled cross-attention alone yields very inaccurate control, possibly because the of several reasons: (i)trainable parameters are too few, (ii) 
% While positional embeddings are not used in the conditioners of ControlNet-based models \cite{musiccontrolnet,hou2024editing}, it turns out that they are critical inductive bias when the trainable parameters are few, likely because the content to be generated (i.e., music signals) and the conditions of interest (e.g., melody) here are both functions of time.
%, adding positional embeddings turn out to be critical for decoupled cross-attention to harness the control signal. 
Built on the simple idea of combining the rotary positional embeddings (RoPE) \cite{su2024roformer} with decoupled cross-attention, MuseControlLite is a novel framework for controlling time-varying conditions. %Specifically, we employ rotary positional embeddings (RoPE) \cite{su2024roformer}, with a modification that applies rotation to all queries, keys, and values to enhance position-aware attention. 
Our implementation introduces only an additional 8\% trainable parameters relative to the diffusion Transformer backbone~\cite{evans2024stable}.
%enabling the learning of fine-grained control signals while preserving the inference speed of the pretrained model, which lacks such controllability.

\fundwo{Unlike our approach, ControlNet-based models \cite{musiccontrolnet,hou2024editing} incorporate positional embeddings in the self-attention layers of the Transformer blocks. These models process the conditioning input in the latent space—matching the shape of the pre-trained backbone—and reintroduce it into the frozen U-Net decoder \cite{musiccontrolnet} or Transformer blocks \cite{hou2024editing} to guide the generation toward the desired output. We find empirically that 
our approach is more parameter-efficient.}
%this approach is overly parameter-intensive, and that decoupled cross-attention layers~\cite{ye2023ip} offer a more efficient and effective alternative.}
%To reduce computational overhead, we propose to leverage the lightweight, highly controllable decoupled cross-attention mechanism presented in IP-adapter~\cite{ye2023ip}. 

Our model supports multi-attribute control similarly to Music ControlNet~\cite{musiccontrolnet}. 
%Moreover, treating the reference audio signals as another type of time-varying control signals, we also train a separate set of adapters for audio conditioning, enabling audio inpainting and outpainting applications. The two adapter sets are fully compatible and can be combined with the original text condition.
\fundwo{Moreover, by introducing ``audio conditioning,'' MuseControlLite is able to replicate the reference audio in full resolution while providing partial control capability, enabling both audio inpainting and outpainting.} 
Additionally, we adopt multiple classifier-free guidance~\cite{liu2022compositional,brooks2023instructpix2pix} to flexibly regulate the strength of the time-varying conditions, preventing the quality degradation that can result from over-fixation.

%We evaluate our method using the Song Describer dataset~\cite{manco2023thesong} and the same metrics employed by Music ControlNet and Stable Audio Open~\cite{evans2024stable} to extract musical conditions. For the audio inpainting and outpainting tasks, we calculate the novelty curve using a chroma-based self-similarity matrix and analyze its boundaries to illustrate the smooth transitions generated by our approach, even when conditions are unrelated to the original audio content. 

The main contributions are three-fold: \begin{itemize} %\item MuseControlLite: 
\item The first investigation of using positional embeddings for decoupled cross-attention layers in diffusion Transformers for controllable text-to-music generation. 
\item The first \fundwo{fine-tuning method} that handles both attribute and audio control (`text+attribute+audio'). In contrast, existing \fundwo{finetuning methods} only take either attribute control  (`text+attribute') \cite{musiccontrolnet}, or audio control (`text+audio') \cite{tsai2024audio}.
\item Demonstration on the public evaluation benchmark of the Song Describer dataset~\cite{manco2023thesong}, showing that MuseControlLite outperforms existing ControlNet-based approaches~\cite{musiccontrolnet,hou2024editing} in melody control, achieving a 4.5\% improvement in melody accuracy.

%Music ControlNet and Stable Audio Open ControlNet, 
%with additional audio conditioning capabilities.
%for audio outpainting and inpainting, producing smoother transitions and higher-quality results. 
%\item Application of multiple classifier-free guidance techniques to separately handle text, musical attributes, and audio, preventing the quality degradation that can result from over-fixation on any single condition. 
%\item A unified model that handles melody control, rhythm control, dynamics control, as well as audio inpainting and outpainting. 
\end{itemize}

% and AP-adapter\cite{tsai2024audio}

\section{Related Work}
%\subsection{Overview of Controllable Music Generation}
Controllable music generation aims to produce music that aligns with human requirements. On a global scale, controls may include text prompts, tempo (BPM), instrumentation, timbre, or mood. On a more fine-grained or local level, control can be exercised over chords, rhythm, dynamics, melodic lines, and other structural elements, usually as time-varying conditions. We review some existing work below.

\textbf{Training-time control with global conditions.}
One of the most common methods to steer a music generation model is direct fine-tuning. \citet{plitsis2024investigating} explores personalized techniques from image generation, such as DreamBooth \cite{ruiz2023dreambooth} and textual inversion \cite{gal2022image}.
%, for text-to-music tasks. 
%By incorporating a specific concept (e.g., a style or timbre) into a dedicated token, these approaches allow the concept to be invoked in multiple contexts. 
%DreamBooth employs an additional prior preservation loss to mitigate overfitting during the fine-tuning of the entire model. In contrast, textual inversion optimizes only the word embedding—freezing the rest of the model—thereby learning a token that best represents a given concept. 
Mustango \cite{melechovsky2023mustango}, trained with music-focused textual captions, 
learns to 
%further enables the model to
understand instructions about chords, 
%beats, 
tempo, and key. 
Instruct-MusicGen~\cite{zhang2024instruct} fine-tunes MusicGen, enabling music editing with text prompts. MusicGen-style~\cite{rouard2024audio} trains MusicGen from scratch, incorporating a text conditioner and an audio feature extractor to fuse text and audio during inference. 

\textbf{Training-time control with local conditions.}
For more granular controls, MusicGen~\cite{musicgen} prepends a melody-based conditioning tensor (along with text embeddings) and trains the entire model from scratch. Similarly, JASCO~\cite{tal2024joint} appends all conditions to the model’s input across the feature dimension but also requires training from scratch—demanding significant computational resources and making it less flexible to add new conditions. MusiConGen~\cite{musicongen} and Coco-Mulla~\cite{lin2023content} fine-tune MusicGen, enabling chord and rhythm conditions without full retraining. 
Music ControlNet~\cite{musiccontrolnet} uses 
%addresses this issue by using 
zero-initialized convolution layers and a trainable copy as an adapter for fine-tuning.
%allowing the powerful pre-trained model to learn and integrate additional control signals without full retraining.

\textbf{Inference-time optimization.}
Training-free approaches for controllable generation also gain attention for their computational efficiency and adaptability \cite{levy2023controllable,novack2024ditto, kim2024training}. These methods leverage pre-trained models directly, avoiding additional model training. %For instance, 
%DITTO \cite{novack2024ditto2,novack2024ditto} employs a $K$-step optimization procedure that refines an initial latent representation based on user-specified conditions. At each step, it uses the pre-trained diffusion model to generate intermediate results, extracting features from these outputs to iteratively adjust the latent. This method aligns the final result with the desired constraints while preserving the benefits of the pre-trained model. 
However, training-free methods often encounter inherent quality limitations compared to fully trained models. In our work, we address similar objectives with MuseControlLite, which is trained on open-source data and provides functionalities similar to DITTO~\cite{novack2024ditto}.

\section{MuseControlLite}
\subsection{Diffusion Background}
Diffusion models~\cite{ho2020denoising, song2020score} 
%have gained prominence for their robust generative capabilities, particularly in image and audio applications. They 
operate in two stages: a forward process that progressively corrupts a clean sample $\mathbf{x}_0$ over $T$ time steps with noise (forming $\mathbf{x}_1, \dots, \mathbf{x}_T$), and a reverse process that is learned to invert the corruption step by step. Diffusion models are often trained to predict the noise $\boldsymbol{\epsilon}$ added at each time step by minimizing a mean squared error denoising loss.

While early diffusion models use U-Net-like architectures for denoising, 
%music data often appears in discrete form (e.g., MIDI or compressed audio tokens). A 
diffusion Transformers~\cite{peebles2023scalable,xie2024sanaefficienthighresolutionimage} emerge as a more effective way to capture long-range dependencies in data through the attention mechanisms. 
%adapts the denoising process to handle token sequences by using attention mechanisms to capture the long-range dependencies characteristic of musical structures. 
During training, each noisy sequence $\mathbf{x}_t$ is passed through the Transformer along with a time-step embedding to predict $\boldsymbol{\epsilon}$,  $\mathbf{x}_0$ or other intermediate variables~\cite{ salimans2022progressive},  depending on the chosen parameterization. By iteratively refining $\mathbf{x}_t$ through this denoising loop, a coherent musical sequence can be reconstructed. 

In our implementation, we use Stable Audio Open~\cite{evans2024stable}, an open-source text-to-music generation model based on a diffusion Transformer with 24 diffusion blocks. Each block contains both self-attention and cross-attention layers. As will be described later, we modify the cross-attention layers to \fundwo{consider} time-varying conditions.
\subsection{Rotary Position Embedding (ROPE) Background}
Absolute positional embeddings~\cite{radford2019language, clark2020electra} add a learned or sinusoidal vector~\cite{vaswani2017attention} to each Transformer token, and relative positional embeddings~\cite{shaw2018self} incorporate distance offsets between tokens in attention. ROPE \cite{su2024roformer} instead rotates query and key vectors by a position-dependent angle, embedding both absolute and relative information more directly. This is encapsulated by the following equations:
\begin{align}
\mathbf{q}^{T}_{m}\mathbf{k}_{n} &=  \bigl(\mathbf{R}_{\Theta,m}^d\,\mathbf{W}^q\,\mathbf{x}_m\bigr)^\top \bigl(\mathbf{R}_{\Theta,n}^d\,\mathbf{W}^k\,\mathbf{x}_n\bigr) \,,\\
\mathbf{R}_{\Theta, m}^d \mathbf{x} &=
\begin{pmatrix}
x_1~~~~~ \\ x_2~~~~~ \\ \vdots \\ x_{d-1} \\ x_d~~~~~
\end{pmatrix}
%\!\otimes\!
\begin{pmatrix}
\cos m\theta_1 \\ \cos m\theta_1 \\ \vdots \\ \cos m\theta_{d/2} \\ \cos m\theta_{d/2}
\end{pmatrix}
+
\begin{pmatrix}
-x_2 \\ ~~~x_1 \\ \vdots \\ -x_d \\ x_{d-1}
\end{pmatrix}
%\!\otimes\!
\begin{pmatrix}
\sin m\theta_1 \\ \sin m\theta_1 \\ \vdots \\ \sin m\theta_{d/2}\\ \sin m\theta_{d/2}
\end{pmatrix}
\end{align}
%transformer tokens
where \(\mathbf{x}_m\) and \(\mathbf{x}_n\) in Eq.\,(1) are embeddings for tokens at positions \(m\) and \(n\) coming from the last layer, \(\mathbf{q}\) and \(\mathbf{k}\)  the resulting query and key vectors, \(\mathbf{W}^q\) and \(\mathbf{W}^k\) the projection matrices, and \(\mathbf{R}_{\Theta,i}^d\) the rotation matrix with \(\Theta = \left\{\theta_i = 10000^{- \frac{2(i-1)}{d}}, \quad i \in \left[1, 2, \dots, \frac{d}{2} \right] \right\}\), where \(\theta_i\) for each position index \(i\) follows an exponential scaling pattern based on the dimension \(d\).
Note that in Eq.\,(2) we drop the subscript of \(\mathbf{x}\) for simplicity and use $x_j$ to indicate the $j$-th entry ($j \in [1, 2, \dots, d]$) of the vector.

\begin{figure*}[htbp]\label{model structure}
    \centering
    \includegraphics[width=1\textwidth]{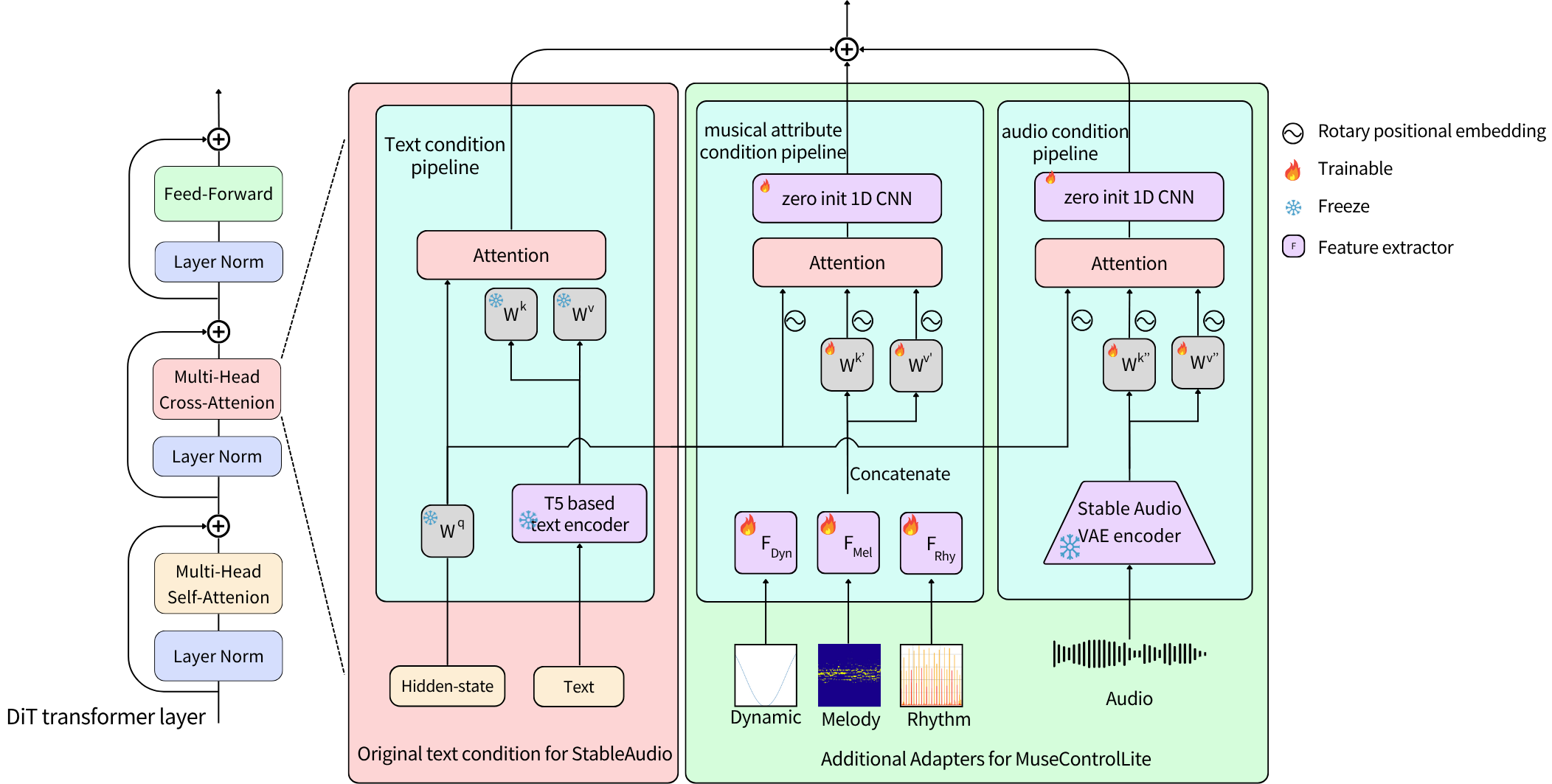} % 調整 width 的比例
    \caption[c]{MuseControlLite uses multiple condition extractors to handle all time-varying controls. The musical attribute offers control over elements such as melody, rhythm, and dynamics, whereas the audio condition enables audio inpainting and outpainting. \fundwo{We train the two pipelines separately with two sets of adapters, but users can choose to use either one or both at the inference time.}}
    %, allowing precise manipulation of these musical features. Meanwhile, t
    %, enabling seamless audio inpainting and outpainting
    \label{fig:MuseControlLite model structure}
\end{figure*}

\subsection{Proposed Adapter Design}\label{subsec:adapter-design}
To prevent over-parametrization yet still achieve the desired fine-tuning performance, Custom Diffusion~\cite{kumari2023multi} demonstrates that fine-tuning only \(\mathbf{W}^k\) (for key vectors) and \(\mathbf{W}^v\) (for value vectors) can suffice for producing personalized outputs with a given new image condition. IP-Adapter~\cite{ye2023ip} further improves this approach by adopting a \textbf{decoupled cross-attention} mechanism that trains only \(\mathbf{W}'^k\) and \(\mathbf{W}'^v\) in the ``decoupled'' layers to learn new conditions related to image generation, where \(\mathbf{W}'^k\) and \(\mathbf{W}'^v\) are learnable copy of (and with parameters initialized from) \(\mathbf{W}^k\) and \(\mathbf{W}^v\), respectively. In view that time-varying conditions in the music/audio domain may be treated similarly as spatial conditions in the image domain, we propose to employ decoupled cross-attention to music generation.

For the original cross-attention layers handling the text condition (i.e., \slwu{the pink-shaded component} in the middle of Figure~\ref{fig:MuseControlLite model structure}), we leave it unchanged:
\begin{equation}\label{eqn:text-cross-attn}
\mathbf{x}_{\text{text}} := \text{Attention}\bigl(\mathbf{x}\,\mathbf{W}^{q},\; \mathbf{c}_{\text{text}}\,\mathbf{W}^{k},\; \mathbf{c}_{\text{text}}\,\mathbf{W}^{v}\bigr)\,,
\end{equation}
where  \(\mathbf{c}_{\text{text}}\) denotes the embedding representing the text condition.

Transferring the technique from the traditional U-Net, we adapt the decoupled cross-attention layers to fit the diffusion Transformer, which calls for additional positional encodings. Specifically, in MuseControlLite, beyond the text condition \(\mathbf{c}_{\text{text}}\), we introduce to the decoupled layers (i.e., \slwu{the musical attribute pipeline within the green-shaded component in Figure~\ref{fig:MuseControlLite model structure}}) an additional musical attribute condition \(\mathbf{c}_{\text{attr}}=\{\mathbf{c}_n\}_{n=1}^N\), which is a function of time with $N$ time points, and \(\mathbf{c}_n\) is at the time position~\(n\) of the sequence \(\mathbf{c}_{\text{attr}}\). 
We apply ROPE to the query, key, and value vectors to enhance the model’s positional awareness. We then train the duplicated \(\mathbf{W}'^{k}\) and \(\mathbf{W}'^{v}\) to handle \(\mathbf{c}_{\text{attr}}\):
\begin{align}
%c_{\text{attr}} & \,,\\
\mathbf{q}_m &= \mathbf{R}_{\Theta,m}^d\,\mathbf{W}^q\,\mathbf{x}_m \,,\\
\mathbf{k}_n &= \mathbf{R}_{\Theta,n}^d\,\mathbf{W}'^k\,\mathbf{c}_n \,,\\
\mathbf{v}_n &= \mathbf{R}_{\Theta,n}^d\,\mathbf{W}'^v\,\mathbf{c}_n \,.
\end{align}
For the decoupled cross-attention, we combine the rotated sequences and calculate the attention:
\begin{equation}\label{eqn:alter-cross-attn}
\mathbf{x}_{\text{attr}} := \text{Attention}\bigl(\mathbf{Q}_{\text{rope}},\, \mathbf{K}_{\text{rope}},\, \mathbf{V}_{\text{rope}}\bigr)\,,
\end{equation}
where $\mathbf{Q}_{\text{rope}}=\{\mathbf{q}_m\}_{m=1}^M$, 
$\mathbf{K}_{\text{rope}}=\{\mathbf{k}_n\}_{n=1}^N$,  
$\mathbf{V}_{\text{rope}}=\{\mathbf{v}_n\}_{n=1}^N$, where $M$ is the length of the musical audio sequence ($N$ is proportional to $M$ but they can be different in general). 
Finally, following~\citet{zhang2023adding}, we add the cross-attention outputs together and connect them with a zero-initialized (zero-init) 1D convolutional layer \(Z_\text{CNN}\) to eliminate initial noise at the start of training. We regard this linear superposition as a correction to the query representation based on the given condition:
\begin{equation}\label{eqn:alter-cross-attn-final}
\mathbf{x} = Z_\text{CNN}(\mathbf{x}_{\text{text}} + \mathbf{x}_{\text{attr}}).
\end{equation}

% As shown in Table~1, introducing ROPE into all key, query, and value vectors in the cross-attention layers triggers the “sudden convergence phenomenon” (noted in ControlNet\,\cite{levy2023controllable}) at an earlier stage. 
\begin{table}[ht] \label{rotary ablation}
    \centering
    \caption{We trained both models for 70,000 steps with a batch size of 32 using the melody condition and found that the one without ROPE struggles to learn the new condition.}
    \label{tab:rope_comparison}
    \begin{tabular}{lcccc}
        \toprule
         & FD${\downarrow}$ & KL${\downarrow}$ & CLAP${\uparrow}$ & Mel acc.${\uparrow}$ \\
        \midrule
        w/o ROPE & 113.13 & 0.58 & \textbf{0.41} & \text{10.7\%} \\
        w/ ROPE & \textbf{~~ 78.50} & \textbf{0.29} & 0.38& \textbf{58.6\%} \\
        \bottomrule
    \end{tabular}
\end{table}

\subsection{Applications for Controls and Manipulations}
We first demonstrate that our model is applicable to all time-varying signals used in Music ControlNet\,\cite{musiccontrolnet}. Next, we incorporate an additional audio condition to enable audio inpainting and outpainting, treating audio signals as another type of control signals.

\paragraph{Musical Attribute Control}
\fundwo{We follow Music ControlNet~\cite{musiccontrolnet} to extract rhythm and dynamics, but adopt a method similar to that of~\citet{hou2024editing} for melody extraction. For melody \(\mathbf{c}_{\text{mel}} \in \mathbb{R}^{N_{\text{melody}} \times 128}\), we first compute the CQT with 128 bins for the mean of the left and right audio channels,\footnote{We noticed that Stable Audio Open ControlNet computes the CQT for the left and right channels separately yields a finer-grained melody condition. Adopting this method in our implementation improved melody accuracy to 64.5\% after only 18,000 training steps.}  apply an argmax operation, then to retain the four most prominent pitches per frame, 
% in each channel
followed by a high-pass filter (cutoff at 261.2\,Hz) to suppress the bass.} Dynamics \(\mathbf{c}_{\text{dyn}} \in \mathbb{R}^{N_{\text{dynamics}} \times 1}\) are derived from the spectrogram energy, converted to decibels, and smoothed with a Savitzky-Golay filter to align with perceived intensity. Rhythm \(\mathbf{c}_{\text{rhy}} \in \mathbb{R}^{N_{\text{rhythm}} \times 2}\) is extracted using a recurrent network-based beat detector~\cite{bock2016madmom} that estimates beat and downbeat probabilities, enabling precise synchronization and rhythmic nuance.

We use separate 1D CNN layers to extract features from each condition and expand their channel dimensions to \( C_r / 3 \), where \( C_r \) is the cross-attention dimension. We then use PyTorch’s \texttt{interpolate} function to match the sequence lengths of these features, setting \( N_{\text{interpolate}} \) equal to the query length \( M \). Finally, we concatenate the features along the last dimension, resulting in the condition representation \( \mathbf{c}_{\text{attr}} \in \mathbb{R}^{N_{\text{interpolate}} \times C_r} \) for the decoupled cross-attention input.

During training, we apply a masking strategy similar to Music ControlNet~\cite{musiccontrolnet}, which randomly masks 10\% to 90\% of the condition, and the masks are independent for the three conditions (i.e., melody, dynamics, rhythm). By using such partial conditioning, we find that the model learns to ``disentangle'' these conditions and can improvise for the unconditioned segments. For example, we can specify music attribute conditions only for the 10--20-second segment, leaving the 0--10-second and 20--47-second segments blank for the model to improvise.

\paragraph{Audio Inpainting and Outpainting}
We directly use the VAE-encoded latent \(\mathbf{x}_0 \in \mathbb{R}^{N_{audio} \times A}\), referred to as the “clean latent” for StableAudio, as \(\mathbf{c}_{\text{audio}}\), where \(N_{\text{audio}}\) is the length of the encoded audio and \(A\) is the number of audio latent channels. Since \(\mathbf{c}_{\text{audio}}\) provides far more information than \(\mathbf{c}_{\text{attr}}\), we found that training with both \(\mathbf{c}_{\text{attr}}\) and \(\mathbf{c}_{\text{audio}}\) simultaneously can cause the model to ignore \(\mathbf{c}_{\text{attr}}\). \fundwo{Thus, the audio condition \(\mathbf{c}_{\text{audio}}\) is trained separately using a distinct set of adapters, while the decoupled cross-attention layers remain the same as those used in the musical attribute conditioning pipeline.} The audio condition is applied with a complementary mask to \(\mathbf{c}_{\text{attr}}\) so that there is no overlap between \(\mathbf{c}_{\text{attr}}\) and \(\mathbf{c}_{\text{audio}}\).

By applying masks to \(\mathbf{c}_{\text{audio}}\) during training, \(\mathbf{W}''^{k}\) and \(\mathbf{W}''^{v}\) learn not only to reflect the condition \(\mathbf{k}_n\) on the same position \(\mathbf{q}_n\) but also to attend to distant tokens \(\mathbf{k}_m\), as shown in Figure~\ref{fig:attention_map} \yang{in the appendix}. This yields smooth transitions at the boundary where the audio condition is given and where it is masked. MuseControlLite can thus simultaneously achieve partial audio control and music-attribute control. Since the segments controlled by \(\mathbf{c}_{\text{audio}}\) are more rigid, we propose to use musical attribute conditions to flexibly control the masked audio segments.

\subsection{Multiple Classifier-Free Guidance} \label{sec:multi-cfg}
Classifier-free guidance~\cite{ho2022classifier} is an inference-time method that trades off sample quality and diversity in diffusion models. It is often used in text-conditional audio or image generation to improve text adherence. \citet{song2020score} provides a crucial interpretation that each denoising step can be viewed as ascending along \(\nabla_{ x} \log p_{\theta}(\mathbf{x})\), the \textit{score} of \(p_{\theta}(\mathbf{x})\). Additionally, any input condition \(\mathbf{c}\) can be incorporated into a diffusion model by injecting the embeddings of \(\mathbf{y}\) into cross-attention\,\cite{rombach2022high}, thus modeling \(p_{\theta}(\mathbf{x} | \mathbf{c})\) (and \(\nabla_{ x} \log p_{\theta}(\mathbf{x} | \mathbf{c})\)). \citet{ho2022classifier} shows that we can train a model by randomly dropping the condition, thereby learning both \(\nabla_{ x} \log p_{\theta}(\mathbf{x})\) and \(\nabla_{ x} \log p_{\theta}(\mathbf{x} | \mathbf{c})\). 
%This procedure enables:
%\begin{align}\label{cfg}
%  \nabla_{x} \log p_{\lambda}(x | c) 
%  &= \nabla_{x} \log p(x)\nonumber \\
%  &\quad + \lambda_{\text{text}}\bigl(\nabla_{x} \log p(x | c) \;-\; \nabla_{x} \log p(x)\bigr).
%\end{align}
Specifically, this procedure enables $\nabla_{ x} \log p_{\lambda}(\mathbf{x} | \mathbf{c}) 
  = \nabla_{ x} \log p(\mathbf{x})  
  + \lambda_{\text{text}}\bigl(\nabla_{ x} \log p(\mathbf{x} | \mathbf{c}) - \nabla_{ x} \log p(\mathbf{x})\bigr)$.
\fundwo{In our work, we use a single pipeline to model \(p_{\theta}(\mathbf{x} | \mathbf{c}_{\text{attr}}, \mathbf{c}_{\text{audio}})\) during training. Our pilot study finds that a fine-tuned model often over-fits to the additional conditions \(\mathbf{c}_{\text{attr}}\) or \(\mathbf{c}_{\text{audio}}\). Therefore, we adopt a separated guidance scale\,\cite{brooks2023instructpix2pix}, \(\lambda_{\text{attr}}\) for musical attributes and \(\lambda_{\text{audio}}\) for audio:}
\begin{equation}\label{eqn:multi-cfg}
\begin{aligned}
&\nabla_{x} \log p_{\lambda}(\mathbf{x} | \mathbf{c}) = \nabla_{x} \log p(\mathbf{x}) \; + \\
&\quad \sum_{i \in \{\text{text}, \text{attr}\, \text{audio}\}} \lambda_i \Bigl(\nabla_{x} \log p(\mathbf{x} | \mathbf{c}_{\leq i}) - \nabla_{x} \log p(\mathbf{x} | \mathbf{c}_{< i})\Bigr).
\end{aligned}
\end{equation}

\fundwo{As detailed in Appendix~\ref{multi_cfg_formulation}, we expand the equation from a single or two conditions~\cite{ho2022classifier,brooks2023instructpix2pix} to a general form of multiple conditions.} 
%Detailed formulation can be found in Appendix~\ref{multi_cfg_formulation}. 

\section{Experimental setup}
\subsection{Dataset}\label{datasets}
For training, we utilize the open-source MTG-Jamendo dataset~\cite{bogdanov2019mtg} and preprocess the data following this pipeline: (i) Resample the audio to 44.1 kHz and segmented into fixed-length clips compatible with the maximum input shape of the Stable Audio Open VAE encoder; (ii) we employ the sound event detection capabilities of PANNs~\cite{kong2020panns} to exclude any audio samples containing vocals; (iii) captions for each audio clip are generated using the Qwen2-Audio-7B-Instruct model~\cite{chu2024qwen2}. Additionally, we remove any samples that overlap with the Song Describer dataset~\cite{manco2023thesong}, as this dataset is reserved exclusively for evaluation purposes. We will also release the code\footnote{\url{https://github.com/fundwotsai2001/Text-to-music-dataset-preparation}} for this data processing pipeline, aiming to facilitate and accelerate future developments in text-to-music model training and fine-tuning.
For evaluation, we adopt the methodology outlined by~\citet{evans2024fast, evans2024stable, evans2024long}. Specifically, we utilize the instrumental subset of the Song Describer dataset, explicitly excluding any prompts associated with vocals. This yields an evaluation set comprising 586 audio clips. All 586 clips are used in the melody-conditioned generation experiments described in Section~\ref{melody compare}, aligning with the settings in Stable Audio Open ControlNet~\cite{hou2024editing}.
\fundwo{However, using musical attributes and text captions from the same audio does not reflect real-world use cases, as users may wish to generate music conditioned on a melody using arbitrary text prompts, rather than the exact caption of the melody source—a process known as \textit{style transfer}. To demonstrate the style transfer capability of MuseControlLite, the tasks described in Section~\ref{ablation} employ a different setup. Ideally, we would have followed the same approach in Section~\ref{melody compare}, but this was not possible due to the closed-source nature of Stable Audio Open ControlNet~\cite{hou2024editing}. To enable style transfer evaluation, we split the 586 audio clips into two disjoint subsets and generated samples by pairing text prompts from the first subset with musical attributes extracted from the second, ensuring that the attributes were independent of the ground-truth audio. The generated outputs were then evaluated against the first subset as the reference. For the audio inpainting and outpainting tasks in Section~\ref{infilling}, we randomly selected a smaller subset of 50 clips from the original 586, without applying the style transfer setting. In Section~\ref{Subjective}, to compare with Stable Audio Open ControlNet, which is open-source, we downloaded example outputs from their project website\footnote{\url{https://stable-audio-control.github.io/}} and generated corresponding samples using the same melody condition audio and text prompts as used by other baselines.
}
\subsection{Training and Inference Specifics}\label{training_inference}
We fine-tune the model using the following objective:
\begin{equation}
\mathbb{E}_{t \sim [0,1], x_t} \left\| f_{\theta} \left( \alpha_t \mathbf{x}_0 + \sigma_t \boldsymbol{\epsilon}, t \right) - \mathbf{v_{t}} \right\|_2^2\,,
\end{equation}
with the velocity term
$\mathbf{v_{t}} = \alpha_t \boldsymbol{\epsilon} + \beta_t \mathbf{x}_0$
following the v-prediction parameterization~\cite{ salimans2022progressive} to improve training stability, where $\alpha_t$ and $\beta_t$ are time-dependent coefficients that balance the contributions of noise and clean data. That is, rather than predicting either the noise or the clean data directly, this parameterization predicts the intermediate variable $\mathbf{v_{t}}$. %$\mathbf{v_{t}}$ defined \begin{equation} \label{v-pred}
%    $\mathbf{v_{t}} = \alpha_t \boldsymbol{\epsilon} + \beta_t \mathbf{x}_0$,
%\end{equation}
%the velocity term \( v_t \) is derived from Equation~\eqref{v-pred}. 
The time variable \( t \) 
is sampled from a noise schedule within the interval \( t \sim [0,1] \). The scaling 
functions are defined as \( \alpha_t = \cos(0.5\pi t) \) and \( \sigma_t = \sin(0.5\pi t) \).

\fundwo{We freeze all model components except for the adapters, feature extractors, and the zero-initialized 1D convolution layers (as shown in Figure~\ref{fig:MuseControlLite model structure}). We train two sets of adapters: one conditioned on all musical attributes and the other conditioned only on audio, using identical training configurations. We use a batch size of 128, a constant learning rate of \(10^{-4}\), and a weight decay of \(10^{-2}\). To encourage the model to focus on \( \mathbf{c}_{\text{attr}} \) or \( \mathbf{c}_{\text{audio}} \), we drop the text condition in 30\% of training iterations. Additionally, each condition is independently dropped with a probability of 50\% and subjected to random masking. The model is trained for 40{,}000 steps with an effective batch size of 128 on a single NVIDIA RTX 3090.
For inference, we fix the separate guidance scales as shown in Table~\ref{tab:guidance_comparison}. We use 50 denoising steps to generate 47-second audio clips. When both a musical attribute condition and an audio condition are applied, we use a complementary masking strategy. This means that the model is exposed to only one type of condition at a time, as the audio condition can be overly dominant and may cause the model to ignore the musical attribute condition.}
\begin{table}[t]
\captionsetup{font=small, skip=3pt} % Set caption size and spacing
\caption{Separated guidance scales used in different tasks}
\label{tab:guidance_comparison}
\centering
\begin{tabular}{@{}lccc@{}}
\toprule
\textbf{Guidance} & \(\lambda_{\text{text}}\) & \(\lambda_{\text{attr}}\) & \(\lambda_{\text{audio}}\) \\
\midrule
Musical attribute control   &7.0&2.0&\ding{55} \\
Audio inpainting              & 7.0                       & 2.0 & 1.0                              \\
Audio outpainting             & 7.0                       & 2.0 & 1.0                            \\ \bottomrule
\end{tabular}
\end{table}

\begin{table*}[h]
\centering
\caption{Melody control comparing with state-of-the-art controllable text-to-music generation models. The proposed model achieves the best melody accuracy and acceptable musical quality metrics with fewer trainable parameters and training data.}
\resizebox{\textwidth}{!}{%
\begin{tabular}{l|ccc|cccc}
\toprule
\textbf{Model} & \textbf{\makecell{Trainable \\ Parameters}} & \textbf{Total Parameters} & \textbf{Training Data} & \textbf{FD ${\downarrow}$} & \textbf{KL ${\downarrow}$} & \textbf{CLAP ${\uparrow}$} & \textbf{Mel acc. ${\uparrow}$} \\
\midrule
MusicGen-Stereo-Large-Melody & ~~3.3B & 3.3B & 20K hr & 193.66 & 0.436 & 0.354 & 43.1\% \\
Stable Audio Open ControlNet & \underline{572M} & \underline{1.9B} & 2.2K hr & ~~97.73 & \textbf{0.265} & \textbf{0.396} & 56.6\% \\
Ours (MuseControlLite-Melody) & ~~\textbf{85M} & \textbf{1.4B} & 1.7K hr & ~~\textbf{76.42} & 0.289 & 0.372 & \textbf{61.1\%} \\
Ours (MuseControlLite-Attr) & ~~\textbf{85M} & \textbf{1.4B} & 1.7K hr & ~~\underline{80.79} & \underline{0.271} & \underline{0.373} & \underline{60.6\%} \\
\bottomrule
\end{tabular}%
}
\label{tab:melody_control}
\end{table*}

% Single table* environment to span both columns and keep tables together
\begin{table*}[ht]
\centering
\begin{minipage}{0.95\textwidth} % Slightly less than textwidth to avoid overfull boxes
\centering
\caption{Performance of all combinations of controls using text and  \(\mathbf{c}_{\text{attr}}\) from different subsets of Song Describer Dataset~\cite{manco2023thesong} (style transfer task). Controllability significantly improves when the relevant condition is provided to the model (\Checkmark).}
% First table
\begin{adjustbox}{width=\textwidth}
\begin{tabular}{l|ccc|ccc|ccc}
\toprule
\textbf{} & \textbf{melody} & \textbf{rhythm} & \textbf{dynamics} & \textbf{FD $\downarrow$} & \textbf{KL $\downarrow$} & \textbf{CLAP $\uparrow$} & \textbf{Mel acc. $\uparrow$} & \textbf{Rhy F1 $\uparrow$} & \textbf{Dyn cor. $\uparrow$} \\
\midrule
{Text condition only} & \ding{55} & \ding{55} & \ding{55} & 124.77 & 0.59 & \textbf{0.42} & 0.09 & 0.21 & 0.05 \\
\midrule
\multirow{3}{*}{{Text$+$single musical attribute}} & \Checkmark & \ding{55} & \ding{55} & ~~\textbf{89.89} & 0.54 & 0.28 & 0.60 & 0.76 & 0.66 \\
 & \ding{55} & \Checkmark & \ding{55} & 111.03 & \textbf{0.45} & 0.38 & 0.09 & 0.89 & 0.42 \\
 & \ding{55} & \ding{55} & \Checkmark & 136.40 & 0.54 & 0.39 & 0.09 & 0.30 & 0.92 \\
\midrule
\multirow{3}{*}{{Text$+$double musical attributes}} & \Checkmark & \Checkmark & \ding{55} & ~~90.00 & 0.56 & 0.28 & \textbf{0.61} & 0.89 & 0.73 \\
 & \Checkmark & \ding{55} & \Checkmark & ~~92.73 & 0.55 & 0.28 & 0.60 & 0.78 & 0.94 \\
 & \ding{55} & \Checkmark & \Checkmark & 119.20 & 0.46 & 0.35 & 0.08 & 0.89 & 0.93 \\
\midrule
{Text$+$all musical attributes} & \Checkmark & \Checkmark & \Checkmark & ~~94.50 & 0.55 & 0.28 & \textbf{0.61} & \textbf{0.90} & \textbf{0.95} \\
\bottomrule
\end{tabular}
\end{adjustbox}
\label{style transfer}
\end{minipage}

\vspace{0.3em} % Minimal space between tables

\begin{minipage}{0.95\textwidth}
% Second table
\centering
\caption{Performance of all combinations of controls using text and  \(\mathbf{c}_{\text{attr}}\) from the same subset of Song Describer Dataset~\cite{manco2023thesong} (non-style transfer task).}
\begin{adjustbox}{width=\textwidth}
\begin{tabular}{l|ccc|ccc|ccc}
\toprule
\textbf{} & \textbf{melody} & \textbf{rhythm} & \textbf{dynamics} & \textbf{FD $\downarrow$} & \textbf{KL $\downarrow$} & \textbf{CLAP $\uparrow$} & \textbf{Mel acc. $\uparrow$} & \textbf{Rhy F1 $\uparrow$} & \textbf{Dyn cor. $\uparrow$} \\
\midrule
{Text condition only} & \ding{55} & \ding{55} & \ding{55} & 124.77 & 0.58 & \textbf{0.42} & 0.09 & 0.22 & 0.06 \\
\midrule
\multirow{3}{*}{{Text$+$single musical attribute}} & \Checkmark & \ding{55} & \ding{55} & ~~\textbf{83.89} & 0.31 & 0.37 & \textbf{0.61} & 0.77 & 0.67 \\
 & \ding{55} & \Checkmark & \ding{55} & 104.83 & 0.41 & 0.40 & 0.09 & 0.86 & 0.48 \\
 & \ding{55} & \ding{55} & \Checkmark & 133.43 & 0.52 & 0.40 & 0.09 & 0.32 & 0.92 \\
\midrule
\multirow{3}{*}{{Text$+$double musical attributes}} & \Checkmark & \Checkmark & \ding{55} & ~~86.62 & 0.30 & 0.39 & \textbf{0.61} & \textbf{0.89} & 0.74 \\
 & \Checkmark & \ding{55} & \Checkmark & ~~88.30 & \textbf{0.28} & 0.38 & 0.60 & 0.78 & 0.94 \\
 & \ding{55} & \Checkmark & \Checkmark & 110.60 & 0.36 & 0.40 & 0.09 & 0.88 & 0.93 \\
\midrule
{Text$+$all musical attributes} & \Checkmark & \Checkmark & \Checkmark & ~~90.28 & 0.31 & 0.39 & \textbf{0.61} & \textbf{0.89} & \textbf{0.95} \\
\bottomrule
\end{tabular}
\end{adjustbox}
\label{non-style transfer}
\end{minipage}
\end{table*}

% Text after tables to prevent overlap with subsequent content

\subsection{Baselines}
Although musical attribute control, audio inpainting, and outpainting have been explored in many prior works, they are either not open-source (e.g., Music ControlNet~\cite{musiccontrolnet}, DITTO~\cite{novack2024ditto, novack2024ditto2}, JASCO~\cite{li2024jen}) or generate a relatively short audio~\cite{tal2024joint}. MusiConGen~\cite{musicongen} and Coco-Mulla~\cite{lin2023content} provide rhythm control; however, \textit{MusiConGen} represents rhythm using a constant beats-per-minute (BPM) value, while \textit{Coco-Mulla} relies on a separate drum track for rhythm representation. We consider both approaches unsuitable for direct comparison with our method. \\
\begin{itemize}
    \item \textbf{MusicGen} \cite{musicgen}: MusicGen is a Transformer-based autoregressive model for text-to-music generation. We adopt the MusicGen-Stereo-Large for audio inpainting and outpainting, MusicGen-Stereo-Large-Melody for melody comparison.
    \item \textbf{Stable Audio Open ControlNet} \cite{hou2024editing}: The ControlNet structure, widely used in text-to-image generation, can also be applied for text-to-music control. Although Stable Audio Open ControlNet is not open-source \cite{hou2024editing}, we 
    employed exactly the same metrics and dataset for comparability. In addition, we contacted the authors to ensure that we followed the same method for extracting melodies and calculating accuracy.
    \item \textbf{Na\"ive masking}: An inference-time method used in DITTO~\cite{novack2024ditto}, we implemented it with Stable Audio Open~\citep{evans2024stable} for audio inpainting and outpainting. We initiate the denoising process with random noise \( \mathbf{x}_t \), and after each denoising step, we immediately overwrite the ``reference'' region of \( \mathbf{x}_{t-1} \) with a noisy version of a given reference audio. 
\end{itemize}

\subsection{Objective Evaluation Metrics}
We use the following metrics to evaluate the musical attribute controllability, text adherence, and audio realism. In order to benchmark against Stable Audio Open ControlNet~\cite{hou2024editing}, we use the same open-source code\footnote{\url{https://github.com/Stability-AI/stable-audio-metrics}} for calculating \text{FD$_{\text{openl3}}$}, \text{KL$_{\text{passt}}$}~\cite{koutini2021efficient}, and \text{CLAP$_{\text{score}}$}~\cite{wu2023large}. \text{FD$_{\text{openl3}}$}~\cite{cramer2019look} extends Fréchet Distance (FD) to full-band stereo audio using OpenL3 (48kHz) features, enabling more comprehensive similarity evaluation. \text{KL$_{\text{passt}}$} measures semantic alignment via KL divergence using PaSST, an AudioSet-trained tagger~\cite{gemmeke2017audio}, adapting it for long-form audio by segmenting and averaging logits. CLAP$_\text{score}$~\cite{wu2023large} assesses text-audio correspondence using CLAP embeddings with a feature fusion approach, ensuring robust evaluation of long-form, high-resolution audio.

%Following~\citep{musiccontrolnet} and \citep{hou2024editing}, we use the metrics below to evaluate music attribute controllability.

\paragraph*{Melody Accuracy} 
% Melody accuracy measures the agreement between the frame-wise pitch classes (C, C\#, ..., B; 12 in total) of the input melody control and those extracted from the generated output. A higher accuracy indicates better preservation of the intended melody.
\fundwo{We directly use the code for calculating melody accuracy from Stable Audio Open ControlNet~\cite{hou2024editing}, provided by the authors, to ensure a fair comparison. The chromagram \( C \in \mathbb{R}^{12 \times T} \) is computed via STFT with a window size of 2048 and a hop size of 512, and we use \texttt{argmax} to select the most prominent pitch. We measure the agreement between the frame-wise pitch classes (C, C\#, ..., B; 12 in total) of the reference audio and the generated audio. Higher accuracy indicates better preservation of the intended melody.}

\paragraph*{Dynamics Correlation} 
\yang{Following~\citet{musiccontrolnet},} we compute Pearson’s correlation to measure the strength of the linear relationship between the dynamics curve of the generated audio and the ground truth condition.

\paragraph*{Rhythm F1} 
\yang{Following~\citet{musiccontrolnet},} the Rhythm F1 score is computed following standard beat and downbeat detection evaluation methods~\cite{davies2009evaluation, raffel2014mir_eval, musiccontrolnet}. It measures the alignment between beat and downbeat timestamps estimated from the input rhythm control and those extracted from the generated audio. These timestamps are obtained using a Hidden Markov Model (HMM) post-filter~\cite{krebs2015efficient} applied to frame-wise beat and downbeat probabilities. In accordance with~\cite{raffel2014mir_eval, musiccontrolnet}, an input and generated (down)beat timestamp are considered aligned if their temporal difference is less than 70 milliseconds.

\paragraph*{Smoothness Value}
To evaluate the smoothness of boundaries in audio inpainting and outpainting tasks, we adopt the Novelty-Based Segmentation approach \cite{muller2015fundamentals}. First, we compute a linear spectrogram and construct the self-similarity matrix (SSM) by measuring pairwise similarities between feature vectors at different time frames. The SSM is then convolved with a checkerboard-shaped kernel centered at each diagonal position. This kernel is designed to emphasize local changes by contrasting regions of high and low similarity. We select a kernel size of 3 to detect novelty on a short time scale, with other hyperparameters same as the tutorial\footnote{\url{https://www.audiolabs-erlangen.de/resources/MIR/FMP/C4/C4S4_NoveltySegmentation.html}} of \citet{muller2015fundamentals}. Finally, the convolution values are summed for each time position, forming a one-dimensional curve known as the novelty curve \(V\). This curve represents abrupt changes in musical content over time, with peaks typically indicating segment boundaries. \fundwo{We define the Smoothness Value as the second finite difference of the position \(V_i\), which is \(V_{i-1} - 2V_i + V_{i+1} \), where \( i \) represents an inpainting or outpainting boundary position.}
A lower Smoothness Value implies 
\yang{non-smooth transition}.
%the} \slwu{abundance of abrupt changes in the frames surrounding the boundaries, which we deem undesirable.}

\begin{table*}[htp]
\centering
\caption{Result for audio outpainting task.\label{audio outpainting}}
\begin{adjustbox}{width=\textwidth}
\begin{tabular}{c|l|ccc|ccc|c}
\toprule
&\textbf{Model}&\textbf{FD ${\downarrow}$}&\textbf{KL ${\downarrow}$}&\textbf{CLAP ${\uparrow}$}&\textbf{Mel acc. ${\uparrow}$}&\textbf{Rhy F1 ${\uparrow}$}&\textbf{Dyn cor. ${\uparrow}$}&\textbf{{\makecell{Smoothness \\ (at 24s) ${\uparrow}$}}}\\
\midrule
\multirow{2}*{{Baseline}}&{MusicGen-Stereo-Large-Melody}&207.48&0.38&0.32&0.19&0.71&0.13&--0.18\\
~&{Na\"ive masking}&193.92&0.64&\textbf{0.39}& -0.07&0.23&0.11&--0.40\\
\bottomrule
\multirow{4}*{{Ours}}&{Text$+$Audio}&184.55&0.37&0.34&0.14&0.45&0.14&--0.16\\
~&{Text$+$Audio$+$Melody}&\textbf{138.07}&\textbf{0.27}&0.35&\textbf{0.57}&0.81&0.43&--0.26\\
~&{Text$+$Audio$+$Rhythm}&165.55&0.32&0.35&0.15&\textbf{0.95}&0.26&--0.16\\
~&{Text$+$Audio$+$Dynamics}&189.15&0.32&0.35&0.14&0.55&\textbf{0.77}&~~\textbf{0.10}\\
\bottomrule
\end{tabular}
\end{adjustbox}
\end{table*}

\begin{table*}[htp]
\centering
\caption{Results for audio inpainting task.\label{audio inpainting}}
\begin{adjustbox}{width=\textwidth}
\begin{tabular}{c|l|ccc|ccc|cc}
\toprule
&&\textbf{FD ${\downarrow}$}&\textbf{KL ${\downarrow}$}&\textbf{CLAP ${\uparrow}$}&\textbf{Mel acc. ${\uparrow}$}&\textbf{Rhy F1 ${\uparrow}$}&\textbf{Dyn cor. ${\uparrow}$}&\textbf{\makecell{Smoothness \\  (at 16s) ${\uparrow}$}}&\textbf{\makecell{Smoothness \\  (at 32s) ${\uparrow}$}}\\
\midrule
{Baseline}&{Na\"ive masking}&193.92&0.64&\textbf{0.39}&0.12&0.29&0.07&--0.62&--0.53\\
\midrule
\multirow{4}*{{Ours}}&{Text$+$Audio}&144.23&\textbf{0.26}&0.30&0.17&0.51&0.14&--0.36&--0.30\\
~&{Text$+$Audio$+$Melody}&\textbf{127.00}&0.33&0.31&\textbf{0.57}&0.80&0.41&--0.56&\textbf{--0.13}\\
~&{Text$+$Audio$+$Rhythm}&144.96&\textbf{0.26}&0.31&0.17&\textbf{0.89}&0.21&\textbf{--0.08}&--0.77\\
~&{Text$+$Audio$+$Dynamics}&150.90&0.29&0.29&0.17&0.61&\textbf{0.74}&--0.40&--0.59\\
\bottomrule
\end{tabular}
\end{adjustbox}
\end{table*}

\section{Result}
\subsection{Melody conditioned generation Comparison}\label{melody compare}
\fundwo{To compare melody-conditioned generation with Stable Audio Open ControlNet~\cite{hou2024editing} and MusicGen-Melody~\cite{musicgen}, we train another version of MuseControlLite that learns only from melody conditions. This version, referred to as \textbf{MuseControlLite-Melody}, is trained for 40{,}000 steps with a batch size of 128. Additionally, we report results for adapters trained with all musical attributes but used for melody-only conditioned generation, referred to as \textbf{MuseControlLite-Attr}. The results are presented in Table~\ref{tab:melody_control}.}
Despite having fewer trainable parameters and less training data, both MuseControlLite-Melody and MuseControlLite-Attr outperform other baselines in terms of FD and melody accuracy, demonstrating superior controllability and realism. However, our models perform slightly worse than Stable Audio Open ControlNet~\cite{hou2024editing} on KL and CLAP metrics.
This may be due to our training data \slwu{being} sourced solely from the MTG-Jamendo dataset~\cite{bogdanov2019mtg}, whereas Stable Audio Open ControlNet~\cite{hou2024editing} uses a more diverse dataset consisting of MTG-Jamendo~\cite{bogdanov2019mtg}, FMA~\cite{defferrard2016fma}, MTT~\cite{law2009evaluation}, and Wikimute~\cite{weck2024wikimute}.

\subsection{Ablation Study for Musical Attribute Conditions}\label{ablation}
\fundwo{We evaluate all combinations of controls on both style transfer and non-style transfer tasks using MuseControlLite-Attr, with the results presented in Tables~\ref{style transfer} and \ref{non-style transfer}. While MuseControlLite-Attr is trained on all musical attribute conditions, it allows users to perform inference with any combination of these conditions.The style transfer task uses musical attributes and text from different subsets, while the non-style transfer task uses musical attributes and text from the same subset. The results in Table~\ref{style transfer} are consistent with what reported in Music ControlNet~\cite{musiccontrolnet}. When the condition is provided, the melody accuracy, rhythm F1, and dynamics correlation are significantly higher. Additionally, we found that when only the melody condition is given, the generated audio still achieves a high rhythm F1 and dynamics correlation, suggesting that the melody condition provides rich information that may include rhythm and dynamics cues.
Comparing Tables~\ref{style transfer} and \ref{non-style transfer}, we find that MuseControlLite-Attr performs worse on the style transfer task for FD, KL, and CLAP.
The degradation of CLAP and KL in style transfer when using melody controls may indicate that the melody condition itself includes the timbre or genre information of the original audio.
Hence, the generated audio still retains a certain level of information relevant to the original audio.
When rhythm and/or dynamics controls are applied, the degradation of CLAP and KL is milder.
On the other hand, musical attribute controllability remains approximately the same for both the style transfer and non-transfer tasks.
} 
\subsection{Audio Outpainting and Inpainting}\label{infilling}
\paragraph{Audio Outpainting}
We mask a 47-second audio clip, retaining only the first 24 seconds, and experiment 
\yang{with no attribute controls and single musical attribute controls}
%with unconditional generation (without \(\mathbf{c}_{\text{attr}}\)) and single musical attribute control 
using MuseControlLite \yang{(i.e., the one trained to consider both attribute controls and audio controls)} and other baselines. We trim the first 24 seconds and retain the outpainting parts, as we aim to evaluate only the newly generated segments.
As shown in Table~\ref{audio outpainting}, our text condition only outpainting outperforms MusicGen-Large in all aspects except for Rhythm F1, demonstrating superior audio realism and consistency. Notably, our non-autoregressive model achieves better objective results than the state-of-the-art autoregressive model, despite autoregression being intuitively preferred for continuation tasks. This suggests that our model effectively learns to improvise missing segments using cross-attention layers, even when no audio condition is provided. \fundwo{While the naive masking method achieves better text adherence according to the CLAP score, it has the lowest smoothness value, indicating abrupt changes at the boundaries.}
\paragraph{Audio Inpainting}
To evaluate audio inpainting, we retain only the first and last 5 seconds of the reference audio, testing the model’s ability to fill in the missing middle portion. Similar to the outpainting task, we exclude the reference segments from the evaluation and assess only the generated inpainted part (i.e., the generated audio from 16s to 32s).
The results, presented in Table~\ref{audio inpainting}, show that in terms of audio realism and text adherence, the performance is similar to that of audio outpainting. However, musical attribute control appears to be more challenging. A possible reason is that, in audio inpainting, the model must handle two transitions---one at the beginning and one at the end---rather than simply continuing the sequence, making the task inherently more complicated.

\fundwo{\subsection{Subjective Evaluation}\label{Subjective}}
\fundwo{
The task for subjective evaluation is similar to Section~\ref{melody compare}, but the dataset consists of samples from the demo website of Stable Audio ControlNet~\cite{hou2024editing}. We apply the same text and melody conditions as demonstrated on their site to generate music using both our model and MusicGen. These outputs are then compared to the samples retrieved from their demo page.}
Participants rate each audio on a 5-point Likert scale based on the following three aspects:
\begin{itemize}[leftmargin=*,noitemsep,nolistsep]
    \item \textbf{Text adherence (T)}: Does the generated audio match the text prompt?
    \item \textbf{Melody similarity (M)}: Does the generated audio faithfully preserve the melody from the reference audio?
    \item \textbf{Overall preference (O)}: Overall, how much do you like the generated audio?
\end{itemize}
\begin{table}[t]
\centering
\caption{User study mean opinion scores (1–5) comparing text adherence (T), melody similarity (M), and overall preference (O).}
\label{tab:comparison}
\small
\begin{tabular}{lccc}
\toprule
\textbf{Model} & \textbf{T} & \textbf{M} & \textbf{O} \\
\midrule
MusicGen-Stereo-Large-Melody   & 3.12 & 2.67 & 3.06 \\
Stable Audio Open ControlNet   & \textbf{3.69} & \underline{4.17} & \textbf{3.65} \\
Ours (MuseControlLite)      & \underline{3.58} & \textbf{4.21} & \underline{3.63} \\
\bottomrule
\end{tabular}
\end{table}
The Stable Audio ControlNet~\cite{hou2024editing} demo page provides eight music editing samples. We divided these into four questionnaires, each containing two melody-conditioned audio samples and three audio samples generated by our baseline methods. A total of 34 participants were recruited from our social circle and randomly assigned to one of the four questionnaires.
%The study took approximately 15 minutes to complete.
According to subjective results, MuseControlLite demonstrates performance comparable to that of Stable Audio Open ControlNet~\cite{hou2024editing} while using fewer training parameters and fewer training data. 

\section{Limitations}
Although MuseControlLite has demonstrated superior performance in controllable music generation, there are still a few weaknesses:
(i) Using multiple classifier-free guidance mechanisms for flexibility slightly slows inference due to multi-batch processing.
(ii) For audio inpainting and outpainting, if the text prompt deviates significantly from the reference audio, MuseControlLite struggles to generate smooth transitions. (iii) MuseControlLite is trained solely on the public MTG-Jamendo dataset~\cite{bogdanov2019mtg}, which contains mostly electronic music. Therefore, MuseControlLite would perform better on other genres after being fine-tuned with a more diverse dataset.
\section{Conclusion}
In this paper, we have introduced MuseControlLite, a lightweight training method that not only enables precise control over music generation under specified musical attribute conditions, but also supports audio outpainting and inpainting, either through text-only condition generation or with musical attribute control. We benchmark our approach against state-of-the-art ControlNet-based methods~\cite{zhang2023adding, hou2024editing} for structural control. MuseControlLite demonstrates superior results, suggesting that it is a powerful alternative to competing models.

Promising future directions include:
(i) manipulating the attention mechanism for more efficient training and improved control precision, and
(ii) enhancing control over conditions that cannot be accurately extracted using current feature extraction methods.

\section*{Impact Statement}
MuseControlLite lowers the barrier for precise, time-varying control in text-to-music generation, making advanced creative tools more accessible to a broader range of artists, hobbyists, and researchers. By introducing a lightweight fine-tuning mechanism with fewer trainable parameters, we enable resource-constrained developers to integrate powerful controllability features into their systems without requiring large-scale computational infrastructures. This democratization of sophisticated AI-driven music creation can foster new waves of artistic experimentation, support rapid prototyping for commercial applications, and reduce the technical overhead that often limits innovation.

At the same time, these capabilities raise important questions about intellectual property, cultural expression, and the changing role of human creators. While MuseControlLite offers transformative advantages—such as facilitating customized soundtracks, educational tools for music learning, and expanded accessibility for individuals with limited musical training—it also highlights the need for responsible use. We encourage practitioners to adopt transparent data governance and to respect copyright laws and cultural contexts when generating music. By balancing creative freedom with ethical considerations, MuseControlLite has the potential to advance the state of music AI while emphasizing responsible and respectful innovation.

\section*{Acknowledgment}

This work is supported by grants from Google Asia Pacific and the National Science and Technology Council of Taiwan (NSTC 112-2222-E002-005-MY2, NSTC 114-2124-M-002-003, NSTC 113-2628-E-002-029, NTU-113V1904-5). The authors sincerely appreciate the valuable feedback provided by the anonymous reviewers. We also thank Prof. Chris Donahue from Carnegie Mellon University for providing insights for this work. Additionally, we are grateful to the authors of \citet{hou2024editing} for helping us align our training and evaluation settings.


\bibliography{example_paper}
\bibliographystyle{icml2025}

%%%%%%%%%%%%%%%%%%%%%%%%%%%%%%%%%%%%%%%%%%%%%%%%%%%%%%%%%%%%%%%%%%%%%%%%%%%%%%%
%%%%%%%%%%%%%%%%%%%%%%%%%%%%%%%%%%%%%%%%%%%%%%%%%%%%%%%%%%%%%%%%%%%%%%%%%%%%%%%
% APPENDIX
%%%%%%%%%%%%%%%%%%%%%%%%%%%%%%%%%%%%%%%%%%%%%%%%%%%%%%%%%%%%%%%%%%%%%%%%%%%%%%%
%%%%%%%%%%%%%%%%%%%%%%%%%%%%%%%%%%%%%%%%%%%%%%%%%%%%%%%%%%%%%%%%%%%%%%%%%%%%%%%

\newpage
\appendix
\onecolumn
\section{Separated guidance scale formulation}\label{multi_cfg_formulation}
To expand the classifier-free guidance from a single condition to a general form, we start from:
% 1. Chain Rule Factorization
\begin{equation}
    p(x, c_1, \ldots, c_n) 
    = p(x)\,\prod_{i=1}^n p\bigl(c_i | x,\, c_1, \ldots, c_{i-1}\bigr) \,.
\end{equation}

% [Add your description/explanation for Equation (1) here]
We simply apply the Bayes’ rule:
% 2. Conditional Probability
\begin{equation}
    p(x | c_1, \ldots, c_n)
    = \frac{
      p(x)\,\displaystyle \prod_{i=1}^n p\bigl(c_i | x,\, c_1, \ldots, c_{i-1}\bigr)
    }{
      p(c_1, \ldots, c_n)
    }\,.
\end{equation}

% [Add your description/explanation for Equation (2) here]
Use log scale to convert multiplications to additions:
% 3. Log of the Conditional Probability
\begin{equation}
    \log p(x | c_1, \ldots, c_n)
    = \log(x)
    + \sum_{i=1}^n \log p\bigl(c_i | x,\, c_1, \ldots, c_{i-1}\bigr)
    - \log p(c_1, \ldots, c_n)\,.
\end{equation}

% [Add your description/explanation for Equation (3) here]
Take the derivative to eliminate the constant term:
% 4. Gradient with Respect to z
\begin{equation}
    \nabla_x \log p(x | c_1, \ldots, c_n)
    = \nabla_x \log p(x)
    + \sum_{i=1}^n \nabla_x \log p\bigl(c_i | x,\, c_1, \ldots, c_{i-1}\bigr)\,,
\end{equation}
\begin{equation}
    \sum_{i=1}^n \nabla_x \log p\bigl(c_i | x,\, c_1, \ldots, c_{i-1}\bigr) = \sum_{i=1}^n \nabla_x \log p(\bigl(c_i, x,\, c_1, \ldots, c_{i-1}\bigr)- \log p\bigl(x,\, c_1, \ldots, c_{i-1}\bigr)) \,.
\end{equation}
Then we scale the condition term with guidance  \(\lambda_{i}\). The guidance \(\lambda_{i}\) can be different according to the control strength that is required.
\begin{equation}
    \nabla_x \log p(x | c_1, \ldots, c_n)
    = \nabla_x \log p(x)
    + \sum_{i=1}^n \lambda_{i}\nabla_x  \log p(\bigl(c_i, x,\, c_1, \ldots, c_{i-1}\bigr)- \log p\bigl(x,\, c_1, \ldots, c_{i-1}\bigr)) \,,
\end{equation}
leading to Equation~\eqref{eqn:multi-cfg} shown in Section~\ref{sec:multi-cfg}.

% \section{Results for using different guidance scale formulation}\label{guidance_scale_formulation}
% We have tested the musical attribute pipeline with different \(\lambda_{\text{attr}}\) and found out that there is a trade-off between the original model's ability and the additional control. The line where \(\lambda_{\text{attr}} = 0\), is using the pretrained Stable Audio Open for 30-second audio generation. As the \(\lambda_{\text{attr}}\) goes larger, \textbf{FD} also rises, but \textbf{KL} and \textbf{CLAP} seems to have the optimal between \(\lambda_{\text{attr}} = 1.5\) and \(\lambda_{\text{attr}} = 2.0\). We pick \(\lambda_{\text{attr}} = 2.0\) for the evaluation in Section~\ref{ablation}.
% \begin{table}[h]
% \centering
% \caption{Results for \(\lambda_{\text{attr}}\) comparison}
% \begin{tabular}{c|ccc|ccc}
% \toprule
% {\(\lambda_{\text{attr}}\)} & \textbf{FD ${\downarrow}$} & \textbf{KL ${\downarrow}$} & \textbf{CLAP ${\uparrow}$} & \textbf{Mel acc. ${\uparrow}$} & \textbf{Rhy f1 ${\uparrow}$} & \textbf{Dyn cor. ${\uparrow}$} \\
% \midrule
% {0.0}    & \textbf{113.73} & 0.51 & 0.39 & 0.10 & 0.22 & 0.16 \\
% {1.5}  & 128.29 & \textbf{0.28} & \textbf{0.40} & 0.66 & \textbf{0.87} & 0.94 \\
% {2.0}  & 134.53 & \textbf{0.28} & 0.39 & 0.70 & \textbf{0.87} & \textbf{0.95} \\
% {2.5}  & 146.07 & 0.31 & 0.35 & \textbf{0.72} & \textbf{0.87} & \textbf{0.95} \\
% {3.0}  & 157.65 & 0.34 & 0.36 & \textbf{0.72} & \textbf{0.87} & \textbf{0.95} \\
% \bottomrule
% \end{tabular}
% \end{table}

% \section{Controlling chord and piano roll}\label{Other time-varying conditions}

\section{Ablation study for all key module in MuseControlLite}\label{module ablation}
We conducted an ablation study on the key modules of our proposed method. Without rotary positional embeddings (RoPE) \cite{su2024roformer}, the model cannot perform melody control, although it still achieves the highest CLAP \cite{elizalde2023clap} score. Removing the 1D-CNN condition extractor causes all metrics to drop, and omitting the zero-initialized 1D-CNN layers used to sum cross-attention outputs similarly degrades performance. Finally, doubling the number of attention heads by scaling ${W}'^k$ and ${W}'^v$ yields no improvement compared to the base configuration. 
\begin{table*}[htp]
\centering
\caption{Performance of all combinations of controls using conditions extracted from Song Describer Dataset~\cite{manco2023thesong}.}
\begin{adjustbox}{width=\textwidth}
\begin{tabular}{l|cccc|ccc|c}
\toprule
\textbf{}&\textbf{Extractor}&\textbf{ROPE}&\textbf{Scale up}&\textbf{CNN}&\textbf{FD ${\downarrow}$}&\textbf{KL ${\downarrow}$}&\textbf{CLAP ${\uparrow}$}&\textbf{Mel acc. ${\uparrow}$}\\
\midrule{Ours}~&\Checkmark&\Checkmark&\ding{55}&\Checkmark& \textbf{~~78.50}&\textbf{0.29}&0.38&\textbf{58.6}\%\\
\midrule
{Ours w/o ROPE}
~&\Checkmark&\ding{55}&\ding{55}&\Checkmark&113.13&0.57&\textbf{0.41}&10.7\%\\
{Ours w/o Extractor}
~&\ding{55}&\Checkmark&\ding{55}&\Checkmark&~~94.50&0.33&0.38&57.2\%\\
{Ours w/o CNN}
~&\Checkmark&\Checkmark&\ding{55}&\ding{55}&~~93.30&0.30&0.37&56.7\%\\
{Ours w/ double heads}
~&\Checkmark&\Checkmark&\Checkmark&\Checkmark&~~80.25&\textbf{0.29}&0.38&58.4\%\\

% \XSolid
% \Checkmark
\bottomrule
\end{tabular}
\end{adjustbox}

\end{table*}
% \begin{table*}[htp]
% \centering
% \caption{Performance of all combinations of controls using conditions extracted from Song Describer Dataset~\cite{manco2023thesong}.}
% \begin{adjustbox}{width=\textwidth}
% \begin{tabular}{l|ccc|ccc|ccc}
% \toprule
% \textbf{}&\textbf{melody}&\textbf{rhythm}&\textbf{dynamics}&\textbf{FD ${\downarrow}$}&\textbf{KL ${\downarrow}$}&\textbf{CLAP ${\uparrow}$}&\textbf{Mel acc. ${\uparrow}$}&\textbf{Rhy F1 ${\uparrow}$}&\textbf{Dyn cor. ${\uparrow}$}\\
% \midrule
% {No condition}&\ding{55}&\ding{55}&\ding{55}&122.09&0.61&\textbf{0.42}&0.09&0.22&0.06\\
% \midrule
% \multirow{3}*{{Single condition}}&\Checkmark&\ding{55}&\ding{55}&\textbf{80.79}&0.27&0.37&\textbf{0.61}&0.77&0.69 \\
% ~&\ding{55}&\Checkmark&\ding{55}&100.78&0.37&0.40&0.09&0.87&0.47\\
% ~&\ding{55}&\ding{55}&\Checkmark&133.23&0.54&0.40&0.09&0.32&0.92\\
% \midrule
% \multirow{3}*{{Double conditions}}&\Checkmark&\Checkmark&\ding{55}&83.55&0.25&0.39&\textbf{0.61}&0.88&0.75\\
% ~&\Checkmark&\ding{55}&\Checkmark&84.39&0.24&0.38&\textbf{0.61}&0.82&\textbf{0.95}\\
% ~&\ding{55}&\Checkmark&\Checkmark&107.59&0.32&0.39&0.09&\textbf{0.89}&0.93\\
% \midrule
% {All conditions}&\Checkmark&\Checkmark&\Checkmark&86.78&\textbf{0.23}&0.38&\textbf{0.61}&\textbf{0.89}&\textbf{0.95}\\
% % \XSolid
% % \Checkmark
% \bottomrule
% \end{tabular}
% \end{adjustbox}
% \end{table*}
\begin{figure}[t]\label{attention_map}
\vskip 0.2in
\begin{center}
\centerline{\includegraphics[width=.6\columnwidth]{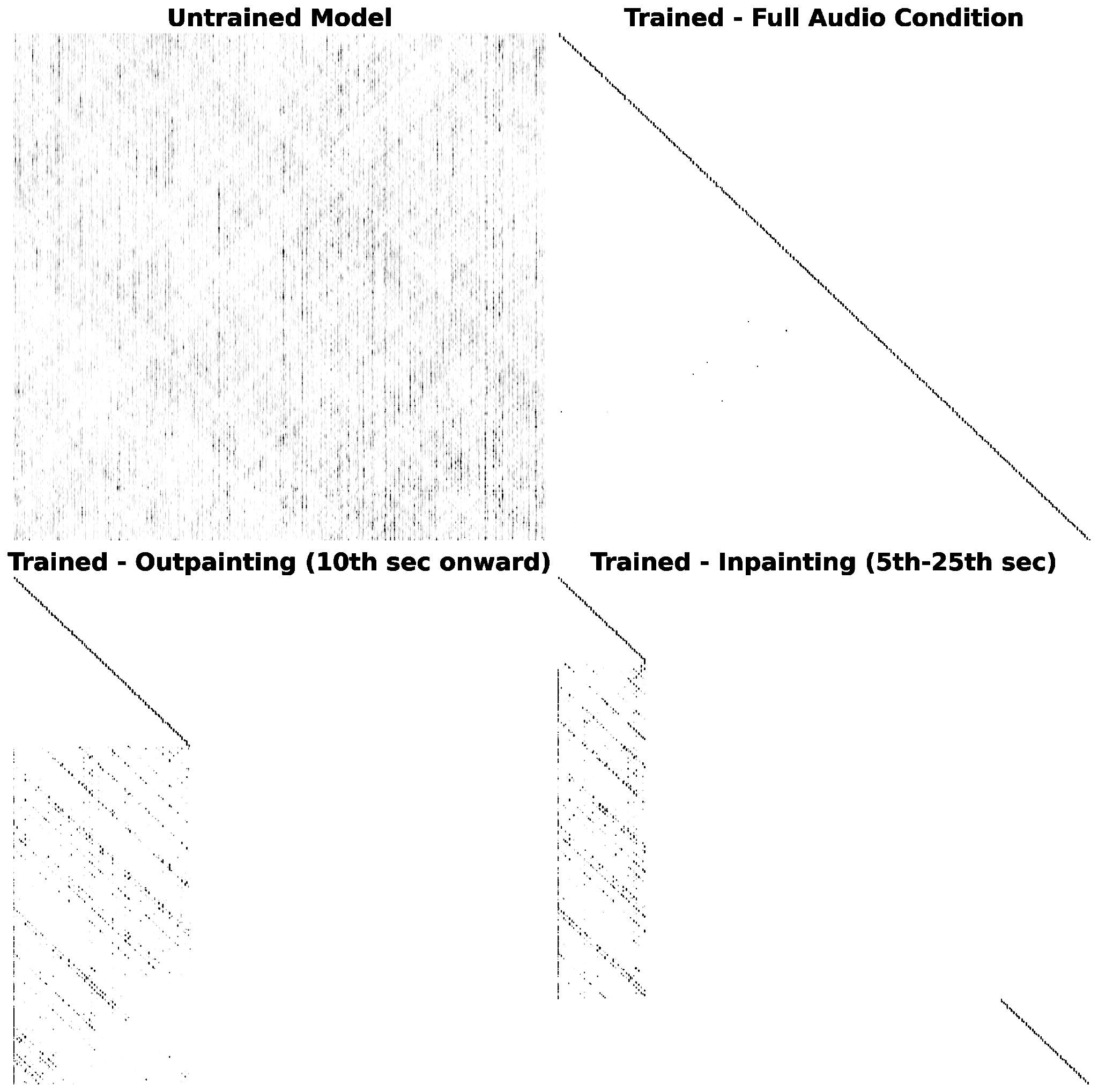}}

\caption{This figure consists of four attention maps: i) untrained, ii) trained, given full audio condition, iii) trained, inpainting 5th-25th seconds, and iv) trained, outpainting 10th seconds onward. After training, a portion of the attention maps exhibits a perfectly diagonal pattern when given the full audio condition.
When performing inpainting or outpainting (i.e., only partial audio condition is given), the model tends to reference previous keys, exhibiting effective usage of given audio context.}
\label{fig:attention_map}

\end{center}
\vskip -0.2in
\end{figure}

\end{document}